\documentclass[prd,showpacs,eqsecnum,nofootinbib,twocolumn]{revtex4}
\usepackage{bm}
\usepackage{stmaryrd}
\usepackage{amsmath}
\usepackage{amssymb}
\usepackage{graphicx}
\usepackage{textcomp}
\usepackage{calrsfs}
\usepackage{yfonts}
\usepackage{epsfig}
\usepackage{pstricks, pst-node, pst-text, pst-3d,pst-coil, epsfig,rotating}
\usepackage{amsmath,graphicx}

\global\arraycolsep=2pt
\newcommand{\ben}{\begin{equation*}}
\newcommand{\een}{\end{equation*}}
\newcommand{\bean}{\begin{eqnarray*}}
\newcommand{\eean}{\end{eqnarray*}}

\newcommand{\nn}{\nonumber}
\newcommand{\be}{\begin{equation}}
\newcommand{\ee}{\end{equation}}
\newcommand{\bea}{\begin{eqnarray}}
\newcommand{\eea}{\end{eqnarray}}
\newcommand{\rmd}{d}
\newcommand{\rme}{e}
\newcommand{\rmi}{i}

\begin{document}
\title{Thermal Issues in Casimir Forces Between Conductors and
Semiconductors}

\author{K. A. Milton}\email{milton@nhn.ou.edu}

\affiliation{
H. L. Dodge Department of Physics and Astronomy, University of Oklahoma,
Norman, OK 73019}

\author{Iver Brevik}\email{iver.h.brevik@ntnu.no}
\author{Simen \AA. Ellingsen}\email{simen.a.ellingsen@ntnu.no}
\affiliation{Department of Energy and Process Engineering, Norwegian University
of Science and Technology, N-7491 Trondheim, Norway}


\begin{abstract}
The Casimir effect between metal surfaces has now been well-verified at the
few-percent level experimentally.  However, the temperature dependence has
never been observed in the laboratory, since all experiments are conducted
at room temperature.  The temperature dependence for the related 
Casimir-Polder force between an atom and a bulk material
 has, in contrast, been observed between a BEC and a silica
substrate, with the environment and the silica held at different temperatures.
There is a controversy about the temperature dependence for the force between
metals, having to do with the magnitude of the linear temperature term
for both low and high temperature, the latter being most prominent at large
distances. There are also related anomalies pertaining to semiconductors.
 The status of this controversy, and of the relevant experiments,
are reviewed in this report.
\end{abstract}
\maketitle
\section{Introduction}

The Casimir effect, reflecting quantum vacuum fluctuations in the 
electromagnetic field in a region with material boundaries, has been studied
both theoretically and experimentally since 1948 \cite{casimir}.  
The forces between 
dielectric and metallic surfaces both plane and curved have been measured at
the 10 to 1 percent level in a variety of room-temperature experiments, 
and remarkable agreement with the zero-temperature 
theory has been achieved. For reviews see 
\cite{Bordag:2009zz,Milton:2004ya,Milton:2001yy}.

In fitting the data various corrections due to surface roughness, patch
potentials, curvature, and temperature have been incorporated.  
It is the temperature correction that is the subject of the present paper.  
Temperature dependence has been 
detected for the Casimir-Polder force between atoms in a
Bose-Einstein condensate \cite{Harber:2005ic,cornell}, the theory of which
was worked out in \cite{Antezza:2008zz}.   A recent experiment by the Yale
group at large distance (0.7--7 $\mu$m) \cite{sushkov}
shows the theoretically expected
reduction of the 
high-$T$ effect for the Casimir force between metallic surfaces.
See also \cite{fth}.

  Theoretically, there are subtle 
issues concerning thermodynamics and electrodynamics which have resulted in
disparate predictions concerning the nature of these corrections.
(An overview is given in \cite{brevik06}.)
However, a general consensus has emerged that the
low-temperature correction to the Casimir effect is relatively large, and that
the linear high-temperature effect should be reduced from the naive expectation
by a factor of 1/2, in accord with the recent Yale  experiment \cite{sushkov}.
(For a critique of this experiment, see \cite{KBM}.)

\section{Conventional approach}

\begin{figure}
 \centering
\includegraphics[scale=0.9]{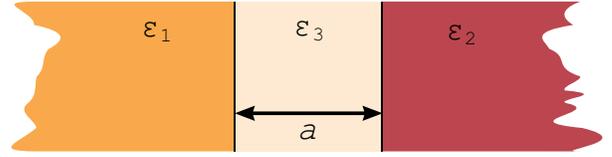}
\caption{\label{figslab} Geometry of two semi-infinite
parallel dielectric slabs, 1 and 2, separated by a third dielectric
slab of thickness $a$.  Each medium is characterized by a permittivity, 
$\varepsilon_i$, which is a function of the frequency.}
\end{figure}

The zero-temperature Casimir effect between parallel conducting plates,
or between parallel dielectrics, is very well understood, and is not 
controversial.  The formula for the latter, which includes the former as a
singular limit, may be derived by a multitude of formalisms 
\cite{dzy,sdm,Bordag:2009zz,Milton:2004ya,Milton:2001yy}. 
For a system of parallel isotropic dielectric media,
as shown in Fig.~\ref{figslab}, characterized by a permittivity
($\varepsilon=\varepsilon(\omega)$)
\be
\varepsilon(z)=\left\{\begin{array}{cc}
\varepsilon_1,&z<0,\\
\varepsilon_3,&0<z<a,\\
\varepsilon_2,&a<z,
\end{array}\right.
\ee
the 
Lifshitz force per unit area on one of the surfaces is at zero temperature
\be
P^{T=0}=-\frac1{4\pi^2}\int_0^\infty \rmd \zeta\int_0^\infty \rmd k_\perp^2
\kappa_3(d^{-1}+d^{\prime -1}),\label{lifshitz0t}
\ee
where $\zeta$ is the imaginary frequency, $\zeta=-\rmi\omega$, and the
longitudinal wavenumber is 
\be
\kappa_i=\sqrt{k_\perp^2+\zeta^2\varepsilon_i(\rmi \zeta)},
\ee
while the transverse electric (TE) and transverse magnetic (TM) Green's 
functions are characterized by the denominators
\begin{subequations}
\bea
d&=&\frac{\kappa_3+\kappa_1}{\kappa_3-\kappa_1}\frac{\kappa_3+\kappa_2}{\kappa_3
-\kappa_2}\rme^{2\kappa_3 a}-1,\\
d'&=&\frac{\kappa'_3+\kappa'_1}{\kappa'_3-\kappa'_1}\frac{\kappa'_3+\kappa'_2}
{\kappa'_3-\kappa'_2}\rme^{2\kappa_3 a}-1,\label{denominators}
\eea
\end{subequations}
respectively, where $\kappa_i'=\kappa_i/\varepsilon_i$.

The attractive Casimir pressure between parallel perfectly conducting planes 
separated by a vacuum space of thickness
$a$ is obtained by setting $\varepsilon_{1,2}\to\infty$ and $\epsilon_3=1$.
In that case the TE and TM contributions are equal, and we have
\be
P_C=-\frac1{8\pi^2}\int_0^\infty \rmd \zeta
\int_{\zeta^2}^\infty \rmd \kappa^2\frac{4\kappa}{\rme^{2\kappa a}-1}
=-\frac{\pi^2\hbar c}{240 a^4},\label{caszero}
\ee
which is Casimir's celebrated result \cite{casimir}.

The controversy surrounds the question of how to incorporate thermal 
corrections into the latter result.  At first glance, the procedure to do
this seems straightforward.  It is well-known that 
thermal Green's functions
must be periodic in imaginary time, with period $\beta=1/T$
\cite{martinschwinger}.  
This implies a Fourier series decomposition, rather 
than
a Fourier transform, where in place of the imaginary frequency integral
above we have a sum over Matsubara frequencies
\be
\zeta_m^2=\frac{4\pi^2 m^2}{\beta^2},
\quad
\int_0^\infty \frac{\rmd\zeta}{2\pi}\to\frac1\beta\sum_{m=0}^\infty{}',
\label{finitetsum}
\ee
the prime being an instruction to count the $m=0$ term in the sum with half
weight.  This prescription leads to the following formula for the Casimir
pressure between perfect conductors at temperature $T$, 
($t=\frac{4\pi a}\beta$)
\be
P^T=-\frac1{4\pi\beta a^3}\sum_{m=0}^\infty{}'\int_{mt}^\infty y^2\,\rmd y
\frac1{\rme^y-1}.\label{perfectt}
\ee

From this it is straightforward to find the high and low temperature
limits,
\begin{subequations}
\bea
P^T&\sim& -\frac1{4\pi\beta a^3}\zeta(3)-\frac{\rme^{-t}}{2\pi\beta a^3}\left(1+t+
\frac{t^2}2\right),\beta\ll 4\pi a,\nn\\ \label{hight}\\
P^T&\sim&-\frac{\pi^2}{240 a^4}\left[1+\frac{16}3\frac{a^4}{\beta^4}
-\frac{240}{\pi}\frac{a}\beta\rme^{-\pi\beta/a}\right],\,
\beta\gg 4\pi a.\nn\\ 
\label{lowt}
\eea
\end{subequations}
These are the results found by Lifshitz \cite{lifshitz}, Fierz \cite{fierz},
Sauer \cite{sauer}, and  Mehra \cite{mehra}.
The two limits are connected by the
duality symmetry found by Brown and Maclay \cite{brown}.
The pressure may be obtained by differentiating the free energy,
\be
P=-\frac\partial{\partial a}F,
\ee
which takes the following form for low temperature (now omitting the 
exponentially small terms)
\be
F\sim -\frac{\pi^2}{720 a^3}-\frac{\zeta(3)}{2\pi}T^3+\frac{\pi^2}{45}
T^4a,\qquad aT\ll1,
\ee
from which the entropy follows,
\be
S\sim-\frac\partial{\partial T}F\sim \frac{3\zeta(3)}{2\pi}T^2-
\frac{4\pi^2}{45}T^3a,\qquad aT\ll1,
\ee
which vanishes as $T$ goes to zero, in accordance with the third law of
thermodynamics, the Nernst heat theorem.

\section{Exclusion of TE zero mode}
\label{sec3}
However, there is something peculiar about the procedure adopted above for a
perfect metal \cite{bostrom}.
 It has to do with the transverse electric mode of zero 
frequency,
which we shall refer to as the TE zero mode.  If we examine the zero frequency
behavior of the reflection coefficients for a dielectric
we see that providing  $\zeta^2\varepsilon(\rmi \zeta)\to0$ 
as $\zeta\to0$, the
longitudinal wavenumber  $\kappa_i\to k$ as $\zeta\to0$,
and hence  $d\to\infty$ as $\zeta\to0$.
This means that there is no TE zero mode for a dielectric.
This statement is not controversial.  
However, if a metal is modeled as the $\varepsilon\to\infty$ limit of a 
dielectric, the same conclusion
would apply.  Because that would spoil the concordance with the third law noted
in the previous section, the prescription was promulgated \cite{lifshitz,sdm}
that the $\varepsilon\to\infty$ limit be taken before the $\zeta\to0$ limit.
But, of course, a real metal is not described by such a mathematical limit, so 
we must examine the physics carefully.

A simple model for the dielectric function is the plasma dispersion relation,
\be
\varepsilon(\omega)=1-\frac{\omega_p^2}{\omega^2},\label{plasma}
\ee
where $\omega_p$ is the plasma frequency.  For this dispersion relation,
the condition $\zeta^2\varepsilon(\rmi\zeta)\to0$ fails to hold as $\zeta\to0$,
and the idealized  prescription result, namely the contribution of the TE
zero mode, follows.

  However, real metals are not well described by this
dispersion relation.  Rather, there is no TE zero mode in the Drude model,
\be
\varepsilon(\rmi \zeta)=1+\frac{\omega_p^2}{\zeta(\zeta+\gamma)},
\label{drude}
\ee
where the relaxation frequency $\gamma$ represents dissipation. The
Drude model  very
accurately fits optical experimental data for the permittivity for
$\zeta  < 2\times 10^{15} $ rad/s \cite{lambrecht1,lambrecht2}.
For gold, appropriate values of the parameters are
$\omega_p=9.03 \mbox{ eV}$, $\gamma=0.0345 \mbox{ eV}$.

Let us review the argument
by writing the Lifshitz formula at finite temperature in the form
\be
P^T=\sum_{m=0}^\infty{}'f_m=\int_0^\infty \rmd m\,f(m)-\sum_{k=0}^\infty
\frac{B_{2k}}{(2k)!}f^{(2k-1)}(0),
\ee
where the second equality uses the Euler-Maclaurin sum formula, in terms
of
\be
f(m)=-\frac1{2\pi\beta}\int_0^\infty \rmd k_\perp^2\,
\kappa(\zeta_m)\left(d_m^{-1}+d_m^{\prime-1}\right),
\label{fm}
\ee
where we assume
that vacuum separates the two plates so $\kappa_3(\zeta_m)=
\kappa(\zeta_m)=\sqrt{k_\perp^2+\zeta_m^2}.$  Here the denominators 
are functions of $\zeta_m$.  By changing the integration variable from
$m$ to $\zeta_m$ we immediately see that the  integral term in the
Euler-Maclaurin sum formula  corresponds precisely to the zero-temperature
result.

One must, however, be careful in computing the low temperature corrections
to this.  One cannot directly expand the denominator $d$ in powers of $\zeta$
because the $k_\perp$ integral ranges down to zero.  Let us 
rewrite the TE term there as follows:
\bea
&&f^{({\rm TE})}(m)=-\frac1{\pi\beta}\int_{2m\pi/\beta}^\infty \rmd \kappa\, 
\kappa^2\nn\\
&&\times
\left\{\left[\frac{1+\sqrt{1+\zeta_m^2(\varepsilon(\rmi\zeta_m)-1)/\kappa^2}}
{1-\sqrt{1+\zeta_m^2(\varepsilon(\rmi\zeta_m)-1)/\kappa^2}}\right]^2
\rme^{2\kappa a}-1\right\}^{-1}.\label{tesummand}
\eea
Evidently, for the Drude model, or more generally, whenever
\be
\lim_{\zeta\to0}\zeta^2[\varepsilon(\rmi \zeta)-1]=0,
\ee
the TE zero mode vanishes,
 $f^{({\rm TE})}(0)=0.$ 
However, it is important to appreciate the physical
discontinuity between $m=0$ and $m=1$ for room temperature.  At 300 K,
while $\zeta_0=0$, $\zeta_1=2\pi T=0.16$ eV, large compared the 
relaxation frequency $\gamma$.  Therefore, for $m>0$,
\bea
&&-\frac1{\pi\beta}\int_{\zeta_m}^\infty \rmd \kappa\,
\kappa^2\left[\left(\frac{\sqrt{1+\omega_p^2/\kappa^2}+1}
{\sqrt{1+\omega_p^2/\kappa^2}-1}\right)^2\rme^{2\kappa a}-1\right]^{-1}
\nonumber\\
&&\qquad\qquad
\approx-\frac1{\pi\beta}\int_{\zeta_m}^\infty \rmd \kappa\,\kappa^2
\frac1{\rme^{2\kappa a}-1},
\eea
provided the significant values of $\zeta_m$ and $\kappa$ are small
compared to the plasma frequency $\omega_p$.  This is just the ideal
metal result.  Insofar as this is accurate,
this expression yields the low- and high-temperature corrections seen above. 
 However, there is now a discontinuity in
the function $f^{({\rm TE})}$. As $\zeta_m\to0$,
\be
f^{({\rm TE})}(m)\to-\frac1{\pi\beta}\int_0^\infty \rmd \kappa\frac{\kappa^2}{
\rme^{2\kappa a}-1}=-\frac{\zeta(3)}{4\pi\beta a^3},
\ee
rather than zero.  

This implies an additional linear term in the pressure
at low temperatures:
\be
P^T\sim P^{T=0}+\frac{\zeta(3)}{8\pi a^3}T, \qquad aT\ll1.\label{linear}
\ee
Exclusion of the TE zero mode will also reduce the linear temperature
dependence expected at high temperatures,
\be
P^T\sim -\frac{\zeta(3)}{8\pi a^3}T,\qquad aT\gg1,\label{hightlinear}
\ee
one-half the usual ideal metal result seen in Eq.~(\ref{hight}).

\section{Proximity Force Approximation}
Most experiments are carried out between a sphere (of radius $R$) and a plane.
In this circumstance, if $R\gg a$, $a$ being the separation between the
sphere and the plate at the closest point, the force may be obtained from
the proximity force approximation (PFA) \cite{derjaguin}, 
\be
\mathcal{F}=2\pi R F(a), 
\ee
$F(a)$ being the free energy for the case of parallel plates separated
by a distance $a$. 
 Thus in the idealized description, the low temperature
dependence including our linear term is for $aT\ll1$,
\be \mathcal{F}\sim
-\frac{\pi^3 R}{360 a^3}\bigg[1-\frac{45}{\pi^3}\zeta(3)aT
+\frac{360}{\pi^3}\zeta(3)(aT)^3-16(aT)^4\bigg].
\ee
Since this conversion is trivial, in the following we will restrict attention
to the straightforward parallel plate situation.
(Recently, use of PFA has been superseded by
numerical/analytical methods, e.g., multiple-scattering 
 and boundary element 
techniques.  See \cite{casimirvol} for a review.)

These results are only approximate, because they assume the metal is ideal 
except for the exclusion of the TE zero.  Elsewhere, we have referred to
this model as the Modified Ideal Metal (MIM) model \cite{brevik03}.
Evidently, for sufficiently low temperatures the
approximation used here, that $\zeta_1\gg\gamma$, breaks down, the 
function $f(m)$ becomes continuous, and the linear term disappears.
Indeed, numerical calculations based on real optical data for the
permittivity show this transition.  
Because this linear behavior does not persist at arbitrarily small 
temperatures,
it is clear that the conflict with the third law anticipated in the
arguments in the previous section do not apply.  
In fact, as we shall see in Sec.~\ref{sec6}, the entropy does go to zero at 
zero temperature for any realistic metal. Only in special idealized model cases 
does the entropy problem persist, wherein $\zeta_1\gg \gamma$ remains true all the 
way to zero temperature; i.e., when $\gamma$ tends to zero as $\gamma\propto T$ or faster 
as $T\to 0$ \cite{ellingsen08,intravaia08}.

\section{Kramers-Kronig relation}

As noted above, there are strong thermodynamic and electrodynamic arguments
in favor of the exclusion of the TE zero mode.  Essentially, the point
is that a realistic physical system can have only one state of lowest
energy.  Electrodynamically, one can start from the Kramers-Kronig relation
that relates the real and imaginary parts of the permittivity, required by
causality, which can be written in the form of a dispersion
relation for the electric susceptibility \cite{embook}
\be
\chi(\omega)=\frac{\omega_p^2}{4\pi}\int_0^\infty \rmd \omega'
\frac{p(\omega')}{\omega^{\prime 2}-(\omega+\rmi \epsilon)^2}.
\ee
If the spectral function $p(\omega')\ge0$ is nonsingular at the origin,
it is easily seen that $\omega^2\chi(\omega)\to0$ as $\omega\to 0$, which
as shown in Sec.~\ref{sec3} implies the 
absence of the TE zero mode.
Conversely, $p(\omega')$ must have a $\delta$-function singularity at
the origin to negate this conclusion.  This would seem implausible for
any but an overly idealized model.  In contrast, in the Drude model the
spectral function is non-singular, becoming singular only in the plasma-model
limit:
\be
p(\omega')=\frac2\pi\frac\gamma{\omega^{\prime2}+\gamma^2}\to2\delta(\omega'),
\qquad\gamma\to0.
\ee

As noted above, use of the plasma model in the reflection coefficients
would lead to the conventional temperature dependence, but this dispersion
relation is inconsistent with real data. 
It may be argued \cite{bezerra04} that in the ideal Bloch-Gr\"uneisen model 
\cite{bloch} the relaxation parameter goes to zero as $\gamma\propto T^5$ near 
zero temperature, thus causing the entropy to tend to a non-zero value at zero 
temperature. Physically this idealized case was shown to correspond to a 
glass-like state of frozen, randomly oriented bulk currents, whose ground 
state is strongly degenerate, hence retaining finite entropy in this limit 
without thereby violating the third law \cite{intravaia09,intravaia10}. 

More to the point, however, real materials exhibit scattering by impurities 
and certainly from boundaries,\footnote{Indeed, for very pure metals $\gamma$ 
is found experimentally to be sample-size dependent, and hence reaches a nonzero 
value even for an imperfection-free metal when its Bloch-Gr\"uneisen electron 
mean free path becomes comparable to sample dimensions \cite{Burns}.}
so that $\gamma$ remains finite. In any case, at sufficiently low temperatures 
the residual value of the relaxation parameter does not play a role,
as the frequency
characteristic of the anomalous skin effect becomes dominant \cite{svetovoy}.
Moreover, the Purdue group \cite{decca} also extrapolate the 
plasma formula from the infrared region down to zero frequency, whereas in fact
frequencies very small compared to the frequency corresponding to the
separation distance play a dominant role in the temperature dependence.

 Finally, we emphasize that all present Casimir experiments
are carried out at room temperature, where the known room temperature data
are relevant.

\section{Third Law of Thermodynamics}\label{sec6}
\begin{figure}
  \begin{center}
    \includegraphics[width=3in]{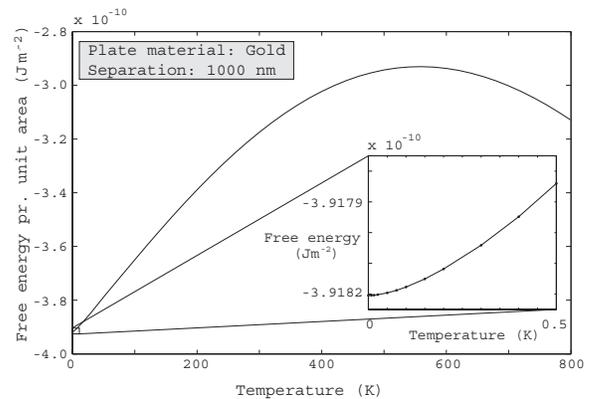}
    \caption{Numerical evaluation of the  free energy between two gold 
    halfspaces as a function of temperature. The inset gives details for low 
    $T$.}
    \label{fig_F(T)}
  \end{center}
\end{figure}

The principal reason for the theoretical controversy has to do with the
purported violation of the third law of thermodynamics if the TE zero mode
is not included.  If ideal metal reflection coefficients are used otherwise
(the MIM model) such a violation indeed occurs, because the free energy
per unit area for small temperature then behaves like
\be
F=F_0+T\frac{\zeta(3)}{16\pi a^2}.  
\ee
However, we and others have shown that for real metals,
the free energy per area has a vanishing slope at the origin. Indeed, in the 
Drude model we have \cite{Milton:2004ya}
\be
F=F_0+T^2 \frac{\omega_p^2}{48\gamma}(2\ln2-1),
\ee
for sufficiently low temperatures. The $T^{5/2}$ correction to this
has been calculated in \cite{Hoye:2007se}.
Figure \ref{fig_F(T)}, from that reference, shows a numerical evaluation of the free energy for gold,
and how the low-temperature linear behavior gives way to a flat slope for very small
temperatures. The exact mathematical requirements for the formal violation of the third 
law are reviewed in \cite{ellingsen09}.

There is, however, an intermediate range
of temperatures where it is expected that the entropy is negative.
We do not believe that this presents a thermodynamic difficulty, and reflects
the fact that the electrodynamic fluctuations being considered represent
only part of the complete physical system \cite{brevik05}.
It was also pointed out by Pitaevskii that the ``Casimir entropy'' is not a 
measure of the system's absolute entropy, but is to be understood as the 
difference between the (positive) entropies of an initial and final state, 
and there is no theorem dictating the sign of such a difference \cite{pitaevskii}.

Independent arguments also lend support to our point of view.
 Jancovici and \v Samaj \cite{Jancovici} and  Buenzli and Martin \cite{Buenzli,Buenzli08}
have examined the Casimir force between ideal-conductor walls with
emphasis on the high-temperature limit.  Not surprisingly, ideal inert
boundary conditions are shown to be inadequate, and fluctuations within
the walls, modeled by the classical 
Debye-H\"uckel theory, determine the
high temperature behavior. The linear in temperature  behavior
of the Casimir force is found to be reduced  by a factor of two
from the behavior predicted by an ideal metal, just as seen above.
  This is precisely the
signal of the omission of the $m=0$ TE mode.  Thus, it is very hard to
see how the corresponding modification of the low-temperature behavior
can be avoided. 

It is just the reduction of this dependence by the factor of 1/2
that is seen in the Yale experiment \cite{sushkov}.

\section{Bohr-van Leeuwen theorem}
Bimonte \cite{Bimonte:2009nf} has suggested that the Bohr-van Leeuwen theorem, which says
that classically at thermal equilibrium electromagnetic fields
decouple from matter, explaining why normal metals do not exhibit
strong diamagnetism,  requires that the TE reflection coefficient
vanishes as zero frequency.  Plasma-like prescriptions are
thus ruled out.  
The Bohr-van Leeuwen theorem does not apply to magnetic materials
or superconductors, since quantum effects dominate there.  Thus experiments on
these materials might reveal quite different behaviors.
Nevertheless, the Bohr-van Leeuwen theorem would seem a more natural criterion 
to turn to than the third law, for systems where magnetic effects are 
negligible at room temperature; the former 
applies in the high temperature limit which is exactly 
where the two Casimir predictions differ the most, 
whereas the latter applies only as the
temperature tends to zero, a situation very far from laboratory conditions 
in any experiment reported to date.

\section{Experimental constraints}
\begin{figure}
\centering
  \includegraphics[width = 3in]
  {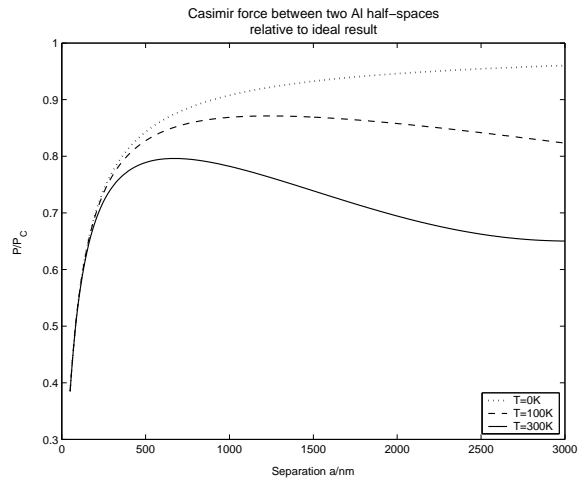}
\caption{\label{alplates} Temperature dependence of the Casimir force between aluminum plates.
The abscissa is the separation in nanometers. }
\end{figure}

We have marshalled theoretical arguments that seem to us quite overwhelming
in favor of the absence of the TE zero mode in the temperature dependence
of the Casimir force between real metal plates, which seem to imply
unambiguously  that there should be large ($\sim 15\%$) thermal corrections to 
the Casimir force at separations of order 1 micrometer.  
Calculations based on this theory, and using real optical data for aluminum, 
for example, have been given, as shown in Fig.~\ref{alplates} for aluminum plates,
taken from \cite{brevik06}.
 The difficulty is that, 
experimentally, it is not easy to perform Casimir force measurements at other 
than room temperature, so 
current constraints on the theory all come from room
temperature experiments. Then all one can do is compare the theory at
room temperature with the experimental results, which must be corrected
for a variety of effects, such as surface roughness, finite conductivity,
and patch potentials.  
A deviation between the zero
temperature theory and room temperature
observations is a measure of the temperature correction.

\begin{figure}
 \centering
\includegraphics[width=2.5in]{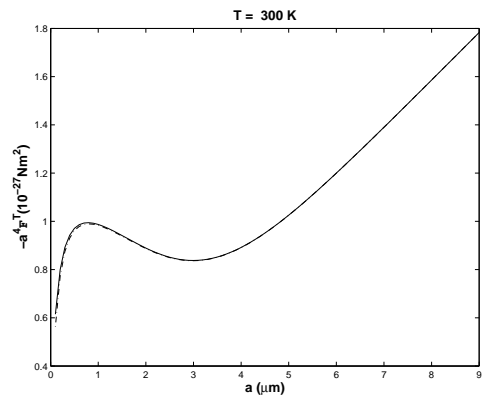}
\caption{The pressure between two gold plates multiplied $a^4$, plotted versus
the plate separation $a$ for room temperature. 
Data for the permittivity is taken from \cite{lambrecht1,lambrecht2}.
This figure is taken from \cite{brevik03}.
\label{fig-fth} }
\end{figure}

Figure \ref{fig-fth} shows the expected behavior of the pressure
as a function of plate separation $a$ at room temperature.  Since,
approximately, the dependence is through the dimensionless factor
(in natural units) $aT$, the high temperature linear behavior is
translated into a linear behavior in the separation.  (The short
distance linear regime is also seen with the expected slope.)

The temperature correction is evidently relatively largest at the largest
separations, where, unfortunately, the total Casimir force is weakest.  
Lamoreaux's new experiment \cite{sushkov} is conducted at the 0.7--7 $\mu$m 
scale,
which is where the linear high-temperature result is dominant.  At these
large distances, the Casimir force is very small, and is overwhelmed
by patch potentials.  Nevertheless, after the patch-potential effect
is extracted and subtracted from the data, the effect is to strongly
favor the Drude-model, that is, exclusion of the TE zero mode, as illustrated
in Fig.~\ref{fig-fth}.
The procedures and  conclusions of \cite{sushkov} are strongly critized in
\cite{KBM}.


It is the most precise experiments of the Purdue group \cite{decca07,decca07a}
 that claim the extraordinarily high precision to
be able to see our effect at distances as small as 160 nm.  Indeed, they
see no deviation from the corrected
zero-temperature Lifshitz theory using optical
permittivities, and hence 
they assert that our theory (the ``Drude model'') is decisively ruled out.
The effect we predict for the temperature correction is only at the 1\% 
level at
a distance of 160 nm, so the measurement must be
performed to that accuracy to see the effect there. (For the usually
employed sphere-plate configuration, $\Delta\mathcal{F}/\mathcal{F}\approx
2.5\%$ at $a=160$ nm.)
Taking the reported accuracy at face value, the nondissipative plasma-model 
approach is favored at a very high confidence level. While doubt has 
been aired 
on a general basis that a 1\%  accuracy can be possible \cite{lamoreaux}, 
we are not aware of any criticism regarding concrete points in Decca's 
impressive experiment, and the results found must be regarded as something of a mystery.


Recently, there has also been a report from the Riverside group
\cite{chang12}, claiming that the Drude model is also disfavored,
although not at as high a level of significance.  We cannot comment
further on this result at present, except to note that the theoretical
support for both the Purdue and the Riverside experiments is identical.

\section{Anomaly for Semiconductors}

Mostepanenko and Klimchitskaya have also claimed there is a similar
anomaly, now affecting the TM reflection coefficient, for dielectrics
and semiconductors \cite{klim05}.
Here is a simple way to understand that argument \cite{ellingsen08p}.
Suppose we model a dielectric with some small conductivity
by the permittivity function
\be
\varepsilon(i\zeta)=1+\frac{\varepsilon_0-1}{1+\zeta^2/\omega_0^2}+\frac{4\pi
\sigma}\zeta.\label{drude-semi}
\ee
(The only essential point is that as $\zeta\to0$, $\varepsilon\to\varepsilon_0$
if $\sigma=0$, otherwise $\varepsilon\to\infty$.)
The Casimir (Lifshitz) pressure is given by
\be
P=-\frac{T}\pi\sum_{i={\rm TE,TM}}
\sum_{m=0}^\infty {}'\int_{\zeta_m}^\infty d\kappa
\,\kappa^2\frac1{(r^{i})^{-2}e^{2\kappa a}-1},
\ee
where the reflection coefficients are
\begin{subequations}
\bea
r^{\rm TE}&=&\frac{-\sqrt{1+\frac{\zeta^2}{\kappa^2}(\varepsilon(i\zeta)-1)}+1}
{\sqrt{1+\frac{\zeta^2}{\kappa^2}(\varepsilon(i\zeta)-1)}+1},\\
r^{\rm TM}&=&\frac{-\sqrt{1+\frac{\zeta^2}{\kappa^2}(\varepsilon(i\zeta)-1)}+
\varepsilon(i\zeta)}
{\sqrt{1+\frac{\zeta^2}{\kappa^2}(\varepsilon(i\zeta)-1)}+\varepsilon(i\zeta)}.
\eea
\end{subequations}

For the case of an ideal metal, it was $r^{\rm TE}$ which was discontinuous:
\be
r^{\rm TE}(\zeta=0)=0, \quad \lim_{\zeta\to0}r^{\rm TE}= -1,
\ee
so this gave a linear temperature term when the sum over Matsubara frequencies
is converted to an integral according to the Euler-Maclaurin formula, for
example.  For a
dielectric the TE reflection coefficient is continuous,
 but if there is
vanishingly small (but not zero) conductivity the TM coefficient is
discontinuous:
\be
r^{\rm TM}(\zeta=0)=1,\quad \lim_{\zeta\to0}r^{\rm TM}=\frac{\varepsilon_0-1}
{\varepsilon_0+1}.
\ee
 This leads to a linear $T$ term in the pressure:
\bea
P_T^{\rm TM}&=&\frac{T}{2\pi}\int_0^\infty d\kappa\,\kappa^2\left[
\frac1{\left(\frac{\varepsilon_0+1}{\varepsilon_0-1}\right)^2 e^{2\kappa a}-1}
-\frac1{e^{2\kappa a}-1}\right]\nonumber\\
&=&\frac{T}{2\pi}\sum_{n=1}^\infty \left[\left(\frac{\varepsilon_0-1}
{\varepsilon_0+1}\right)^{2n}-1\right]\int_0^\infty d\kappa \,\kappa^2 e^{-2
n\kappa a}\nonumber\\
&=&\frac{T}{8\pi a^3}\left[{\rm Li}_3\left(\left[\frac{\varepsilon_0-1}
{\varepsilon_0+1}\right]^2\right)-\zeta(3)\right],
\eea
where the logarithmic integral function is
\be
{\rm Li}_n(\xi)=\sum_{k=1}^\infty \frac{\xi^k}{k^n}.
\ee
Note that the linear term vanishes for $\varepsilon_0\to\infty$.  The free
energy term $F$ is obtained from this by multiplying by $a/2$.  Thus at zero
temperature, the entropy is nonzero,
\be
S=-\left(\frac{\partial F}{\partial T}\right)_V=-\frac1{16\pi a^2}
\left[\mbox{Li}_3\left(\left[\frac{\varepsilon_0-1}{\varepsilon_0+1}\right]^2
\right)-\zeta(3)\right],
\ee
as found in \cite{klim05},
which again appears to violate the third law 
of thermodynamics, stating that the entropy of a nondegenerate system 
must vanish at zero temperature.

We believe that real materials cannot undergo this violation, and that,
properly treated, the entropy must vanish at zero temperature.  The details
of this assertion, however, have not been worked out to our knowledge.
In materials with a very small concentration of free charges, it seems 
highly unlikely that a simple, local, Drude-type model can provide an 
adequate description of reflectivity \cite{ellingsen08}, and 
salvation may well be found to lie in a complete inclusion of spatial 
dispersion effects.

Three simultaneous and very similar efforts were made in this direction 
\cite{pitaevskii08,dalvit,svetovoy08}. These partly seek to bridge the 
gap between the predictions of a naive Drude-like model (\ref{drude-semi}) 
and an experiment performed by Mohideen's group \cite{chen07,chen07a}. 
In that experiment, the carrier concentration of a semiconductor was 
increased by laser illumination, and the observed behaviour of the Casimir 
force is surprising at first sight: 1) When the carrier concentration is high, 
the conductivity must be included if the simple model (\ref{drude-semi}) is used. 
2) When the carrier density is low (but nonzero), the conductivity must be excluded 
in order to reproduce measured data. In light of the spatial dispersion theory 
based on Debye-H\"{u}ckel screening \cite{pitaevskii08,dalvit,svetovoy08,hoye-fc}, 
such behavior is not 
so surprising, however, although these theories fail to reproduce the data perfectly. 
Unlike the mystery of Decca's experiment with metal surfaces, the experimental 
result here appears intuitively reasonable, but a complete theoretical description 
without any ad hoc prescriptions is still lacking.

\section{Real Temperature Experiments}
In any case, it would seem imperative to perform experiments at different
temperatures in order to provide evidence for or against temperature
dependence of Casimir forces. However, such experiments are
extremely difficult, and, for example, Onofrio has abandoned his
attempt to do such a measurement (private communication). 
For example, the increase of temperature from 300 K 
to 350 K should reduce the force by about 2.5\%, for Au-Au plates when the 
separation is 1.0 micro\-meter. 

Even though the new Yale experiment \cite{sushkov}
is consistent with the revised theory, 
it needs to be independently confirmed, especially in view of
the very large electrostatic corrections that must be supplied.
Also, direct measurement of
the temperature dependence would be highly beneficial.  (The only
direct temperature measurement of temperature dependence is that for the
CP force on a BEC of Rb atoms \cite{cornell}.)
 We encourage experimentalists to redouble their efforts,
for the issues involved touch at the heart of our fundamental theoretical
understanding of electrodynamics, statistical mechanics, and quantum
field theory.  

\acknowledgments
The work of KAM was supported by grants from the US National Science
Foundation and the US Department of Energy.  We thank the organizers
of the FQMT11 conference in Prague, Vaclav Spicka, Theo Nieuwenhuizen, 
and Peter Keefe, for putting together such a
successful and interdisciplinary meeting.

\end{document}